\documentclass[useAMS,usenatbib]{mn2e}

\title[Plasma composition]{PSR B1133+16: radio emission height and plasma 
composition}
\author[P. B. Jones]{P. B. 
Jones\thanks{E-mail:p.jones1@physics.ox.ac.uk}\\ 
University of Oxford, Department of Physics, Denys Wilkinson Building,\\
Keble Road, Oxford OX1 3RH, U.K.}

\begin{document}

\date{}

\pagerange{\pageref{firstpage}--\pageref{lastpage}} \pubyear{}

\maketitle

\label{firstpage}

\begin{abstract}

Recent operation of LOFAR by Hassall et al has produced severe 
constraints on the size and altitude of the $40$ MHz emission region in 
this pulsar.  It is shown that these limits, given a limited number of 
unexceptionable assumptions, demonstrate that an electron-positron plasma
cannot be the source of the emission. A physically-acceptable plasma 
source composed of protons and ions arises naturally in pulsars having 
positive corotational polar-cap charge density.  Acceptance of this would 
greatly clarify the classification of pulsar types within the whole.

\end{abstract}


\section{Introduction}

PSR B1133+16 is a small-dispersion-measure pulsar whose radio emission 
exhibits both nulls and sub-pulse drift, but is otherwise unremarkable.  
Its radio brightness has enabled Hassall et al (2012), using LOFAR 
observation in the interval $40$ - $190$ MHz, to give extremely small 
upper limits for the size of the emitting region,
$r_{max} - r_{min} < 4.9\times 10^{6}$ cm.  The maximum distance for the 
$40$ MHz emission is $r_{max} < 1.1\times 10^{7}$ cm from the centre of 
the neutron star, equivalent to $\sim 10^{-3}R_{LC}$, where $R_{LC}$ is 
the light-cylinder radius. The section of the open magnetosphere defined 
by these limits can then be described as a narrow black box, subject to a 
zero-potential condition on the boundary with the closed region, within 
which the plasma mode(s) responsible for the emission couples strongly 
with the radiation field, that is, modes which can propagate in free 
space.

Unfortunately, and for technical reasons, the authors are unable to quote 
confidence levels, but as they note, these results are of some 
significance for the problem of understanding the physics of the emission 
process.  The canonical model for this is a dense plasma of secondary 
electron-positron pairs formed by the conversion of curvature radiation 
(CR) photons or by inverse Compton scattering (ICS) from the primary 
accelerated electrons.  An unstable, possibly quasi-longitudinal, plasma 
O-mode grows in amplitude as the secondary electrons and positrons flow 
outward and transfers energy to radiative modes within the black-box 
volume. The work of Asseo, Pelletier \& Sol (1990) serves as an example 
of such a mode. The growth rate must be large enough to provide the 
necessary gain in amplitude but not so large that the mode is unable to 
accommodate adiabatically to the change in magnetic field strength.

The present paper provides strong evidence that, on the basis of a 
limited number of unremarkable and unexceptionable assumptions, the 
canonical electron-positron pair model is not compatible with the 
observationally-deduced values of $r_{max}$ and of $r_{max} - r_{min}$.

\section[]{The predicted black-box spectrum}

The assumptions about the pair plasma in B1133+16 to which we refer are 
to some degree inter-dependent, but are listed as follows.

(i)		There is a frame of reference, with Lorentz factor $\gamma _{c}$ 
relative to the rotating neutron-star frame, in which the bulk 
electron-positron plasma has zero momentum.

(ii)	The only natural frequencies present in the magnetized plasma 
within the black box are the rest-frame plasma frequency 
$\omega^{c}_{p}$, the critical frequency for curvature radiation $\omega 
_{c}$, and the cyclotron frequency $\omega _{B}$.  At $r_{max}$, the 
transition rate from the electron first Landau excited state to the 
ground is $4\omega^{2}_{B}e^{2}/3mc^{3} \sim 10^{10}$ s$^{-1}$ and 
$\omega _{B}$ is many orders of magnitude greater than any angular 
frequency in the radio spectrum.  Therefore, the plasma is very near the 
high-field limit and does not constitute a complex system likely to have 
unforeseen properties.

(iii)	Coherent curvature radiation can be generated within the black 
box and has both O and E-mode amplitudes.  The critical frequency, above 
which the radiation spectrum cuts off exponentially, is
$\omega _{c} = 3c\gamma^{3}_{c}/2\rho$, where $\rho$ is the flux-line 
radius of curvature.  Whilst propagation of the O-mode is subject to some 
limitations, the E-mode propagates freely in the high-field limit.
The conditions necessary for significant energy transfer to the E-mode 
were described in the original paper of Ruderman \& Sutherland (1975).  
Suitable $\gamma _{c}$ and flux-line radius of curvature may exist in the 
black box to give values of $\omega _{c}$ within the main part of the 
B1133+16 radio spectrum, but a suitable charge density structure is also 
needed to produce coherent emission.  In the plasma rest-frame, this can 
only be associated with $\omega^{c}_{p}$. The radiation wavelength in the 
neutron-star or observer frame is $2\pi c/\omega _{c}$ and there can be 
no effective coherence if, within this interval, there are a large number 
of charge density fluctuations of alternating sign. Thus in the limit 
$\omega _{c} \ll \gamma _{c}\omega^{c}_{p}$
there is no significant energy transfer to either O or E-modes. 

(iv)	The number density of primary accelerated electrons is little 
different from the Goldreich-Julian density and on average, each primary 
electron produces $\lambda$ pairs. This is also not inconsistent with 
recent studies of the force-free magnetosphere using the techniques of 
numerical plasma kinetics (see Chen \& Beloborodov 2013, Timokhin \& 
Arons 2013) in which it is assumed that the open magnetosphere current 
density is determined, principally, by the force-free solution for the 
whole magnetosphere.

(v)	    Outside the black box, the radiative modes do not interact 
strongly with the magnetosphere except that there is the possibility of 
cyclotron resonant absorption in the open magnetosphere region near the 
light cylinder.  Inside the black box, the unstable plasma mode gains 
amplitude until non-linearity, and possibly turbulence, is reached and 
then couples strongly with a distribution of radiative-mode angular 
frequencies.

(vi)	The physical assumptions on which the treatment of aberration and 
retardation by Hassall et al is based are valid.  Thus the energy 
transfer to radiative modes within the black box and in the neutron-star 
frame of reference is to modes with wave-vector locally parallel with the 
magnetic flux density ${\bf B}$.  Cancellation of an aberration and 
retardation time difference, otherwise detectable in principle, by the 
effect of rotation-produced toroidal sweep-back of magnetic flux is not a 
realistic possibility for B1133+16.  The estimate of its size made by 
Shitov (1983), for a vacuum magnetosphere, is dependent on 
$(r/R_{LC})^{3}$ and would be negligible at the radii $r$ found by 
Hassall et al.  Also, these radii lie well inside the spherical inner 
boundary assumed in current numerical studies of the force-free 
magnetosphere. Although field geometry in the real magnetosphere is 
likely to be very different from the vacuum case at $r\sim 10^{-1}R_{LC}$ 
or larger, there is no reason to suppose this could be true at the very 
small radii considered here.

Following these assumptions, the angular frequency of the mode must be 
close to the local plasma frequency which is, in the plasma rest-frame,
\begin{eqnarray}
\omega^{c}_{p} = \left(\frac{8\pi n_{GJ}{\rm e}^{2}\lambda}{m\gamma 
_{c}}\right)^{1/2}
\end{eqnarray}
where $n_{GJ}$ is the Goldreich-Julian number density and $m$ is the 
electron mass. Its Lorentz transformation to either the neutron-star or 
observer frame gives a typical radiation frequency,
\begin{eqnarray}
\nu _{obs} = \gamma _{c}\omega^{c}_{p}/\pi \approx 0.26 \gamma _{c}^{1/2}
\lambda^{1/2} \hspace{5mm}{\rm GHz},
\end{eqnarray}
for B1133+16 when evaluated at $r_{max}$, for the value of the angle 
between ${\bf B}$ and the spin ${\bf \Omega}$ cited by Hassall et al, and 
for a polar surface magnetic flux density of $2.13\times 10^{12}$ G.

The spectrum of B1133+16 is unremarkable (see, for example, Sieber 2002). 
The ATNF pulsar catalogue (Manchester et al 2005) fluxes at $400$ and 
$1400$ MHz give a spectral index $\alpha = -1.65$. A low-frequency 
turnover at $\nu _{max} \approx 60$ MHz has been observed by Deshpande \& 
Radhakrishnan (1992) and above $1400$ MHz, the spectrum steepens to a 
more negative $\alpha$.  The pulsar is observable up to $8.35$ GHz 
(Honnappa et al 2012). Assuming a cut-off at $\nu < \nu _{max}$ and a 
constant $\alpha$ at $\nu > \nu _{max}$, the fraction of the total energy 
flux at frequencies above $\nu$ is $(\nu/\nu _{max})^{1 + \alpha}$.  
Therefore, the frequency below which a fraction $f$ of the total 
radio-frequency spectral energy is contained is,
\begin{eqnarray}
\nu = \nu _{max}(1 - f)^{1/(1 + \alpha)}.
\end{eqnarray}
Thus half the energy  flux is contained below $2.9\nu _{max}$.
Referring back to equation (2), typical cited values of the free 
parameters it contains are $\gamma _{c} = 50$ and $\lambda = 10^{2}$, but 
it must be emphasized that these are either extremely model-dependent or 
are merely estimates that are deemed plausible.  However, for much 
smaller values of $\lambda$, the plasma density becomes too low to allow 
the relative bulk velocity of the electrons and positrons to adjust to 
that local charge density which is needed to give only a very small 
electric field component parallel with ${\bf B}$, and hence maintain only 
small values of $\gamma _{c}$.  Therefore, values $\lambda\gamma _{c} 
\sim 1$ are not a realistic possibility and comparison with equation (2) 
shows that the bulk of the spectral energy of B1133+16  is at frequencies 
far too low to be consistent with that expected for an electron-positron 
plasma. Coherent curvature radiation is not a plausible explanation for 
the reason stated in (iii) above.

Specifically, there seems to be no obvious mechanism for the large-scale 
displacement of spectral energy to frequencies well below those given by 
equation (2).  In fact, we anticipate that the black box has the opposite 
property. Nonlinearity is itself a source of harmonic generation. But 
more significantly, the transfer of energy to a distribution of degrees 
of freedom with higher wavenumbers is a general property of developing 
turbulence (see, for example, Batchelor 1967 for the case of 
incompressible fluids and Weatherall 1997 for the pulsar plasma). It 
occurs through the formation of a region of self-similarity in which the 
energy density as a function of wavenumber
$k$ has a power-law form, $\propto k^{\beta}$.
It is then natural to associate $\nu _{max}$ with the frequencies 
predicted by equation (2), and the power-law high-frequency tail of the 
radio spectrum with a region of self-similarity, although the closeness 
of $\alpha$ for B1133 +16 to the Kolmogorov index $\beta = -1.67$ for 
fluid turbulence must be accidental.  More specifically to the pulsar 
magnetosphere, Weatherall (1997) has demonstrated directly that energy 
transfer to high wavenumbers does occur.

The source of the radio emission proposed here is a plasma of protons and 
ions (mass $M$, mass number $A$, charge $Z$) which is relativistic, but 
with quite low Lorentz factors in the neutron-star frame.  This requires 
the pulsar to have rotational spin such that ${\bf \Omega}\cdot{\bf B} < 
0$.  The details of how this plasma is formed have been described 
elsewhere (Jones 2012, 2013).  Here, it is sufficient to note that 
equation (2) is replaced by,
\begin{eqnarray}
\nu _{obs} = 4.2\left(\gamma _{A,Z}\frac{Z}{A}\right)^{1/2}
\hspace{5mm} {\rm MHz},
\end{eqnarray}
constructed from the ion rest-frame plasma frequency at $r_{max}$,
in which $\gamma _{A,Z}$ is the Lorentz factor of the ion in the 
neutron-star frame.  The plasma is essentially cold: there is a 
substantial difference between the proton and ion Lorentz factors and no 
doubt that the relativistic Penrose condition (Buschauer \& Benford 1977) 
necessary for amplitude gain is satisfied. The anticipated Lorentz factor 
is of the order of $\gamma _{A,Z} \sim 10$ (Jones 2012) which gives a 
frequency of
$\sim 10$ MHz, but the $r_{max}$ found by Hassall et al is an upper limit 
and the ion rest-frame plasma frequency is increased by a factor 
$(r_{max}/r)^{3/2}$  at smaller $r$, so that $\nu _{max} = 60$ MHz for 
B1133+16 is easily attainable. The power-law high-frequency tail at
$\nu > \nu _{max}$ can be identified with the region of self-similarity 
described in the previous paragraph so that the GHz radiation is easily 
understood.
  On the basis that turbulent energy transfer to higher wavenumbers is 
the source of the high-frequency tail in the spectrum, and given that the 
unstable mode itself is quasi-longitudinal, the radiation emission is 
unlikely to be consistent with any simple form of radius-to-frequency 
mapping.

\section{Conclusions}

Rates of self-sustaining pair creation by the magnetic conversion of 
curvature radiation or of ICS photons have been published by Hibschman \& 
Arons (2001) and Harding \& Muslimov (2002, 2011); here we refer to Fig. 
1 of Harding \& Muslimov (2002). For dipole field geometry, the 
constraints on rotation period $P$ and polar magnetic flux density $B$ 
can be expressed in terms of a parameter $X = B_{12}P^{-1.6}$, where 
$B_{12}$ is here the polar magnetic flux density in units of $10^{12}$ G.  
The condition  for CR pair formation is $X > X_{c} \approx 6.5$, and is 
far from being satisfied by B1133+16 for which $X = 1.6$.  A secondary 
pair plasma could be formed only by ICS photons or by higher-multipole 
enhancement of flux-line curvature.  However the condition is indicative, 
not rigorous, because it is always possible that, in a specific pulsar, 
there could be some curvature enhancement.

It is unfortunate that the engagement with electron-positron plasma has 
been so exclusive in pulsar physics because there is no observational 
evidence that $X < X_{c}$ pulsars in general produce pairs. They are so 
old that there are no observable associated nebulae.

Pulsars with $X > X_{c}$ generate very large maximum  acceleration 
potentials which can certainly support CR pair creation for either sign 
of ${\bf \Omega}\cdot{\bf B}$.  Those LAT-detected pulsars found by Abdo 
et al (2010) conform with this class. The detected GeV $\gamma$-rays are 
emitted from the light-cylinder region, presumably from a substantial 
flux of electrons and positrons. Of the 46 pulsars listed in Table 1 of 
Abdo et al, only 7 have $X < X_{c}$ and without exception, these are 
millisecond pulsars.

As $X$ decreases with pulsar age toward $X_{c}$, the density of pair 
plasma decreases until there is a bifurcation in evolutionary path.  In 
the ${\bf \Omega}\cdot{\bf B} > 0$ case, the growth rate of the unstable 
mode eventually becomes too small for observable radio emission.  Those 
with ${\bf \Omega}\cdot{\bf B} < 0$ can continue to produce a proton-ion 
plasma and display phenomena such as nulls and subpulse drift (see Jones 
2012, 2013).

The final paragraph of Jones (2013) suggested that the operation of LOFAR 
might well provide evidence for this latter class of pulsar through  
observations on the shape of low-frequency spectra.  But this is not so.
By direct plotting from the ATNF pulsar catalogue, it can be seen that 
the spectral index $\alpha$ formed from fluxes at $400$ and $1400$ MHz is 
almost uncorrelated with $X$.  Also, for example, low-frequency emission, 
down to $10$ MHz and with no observed turnover, is strong in the Crab 
pulsar (see Sieber 2002).  In fact, it appears that measurements of 
$r_{max}$, where possible, in extensions to the work of Hassall et al 
will be decisive in determining plasma composition.  Whilst the present 
paper is concerned specifically with B1133+16, the remaining three 
pulsars investigated by Hassall et al could have been discussed in the 
same way with broadly similar, though less strong, conclusions.

\section*{Acknowledgments}

The author thanks the referee for some very useful comments.  He first 
became aware of the LOFAR work of Hassall et al at an informal meeting of 
pulsar observers held in Oxford and wishes to thank
Dr Aris Karastergiou for an invitation to attend.

\bsp

\label{lastpage}

\end{document}